\documentclass{article}

\usepackage{fancyheadings}

\usepackage[british]{babel}
\usepackage{amsmath}
\usepackage{amssymb}

\newcommand{\lie}{\mathcal{L}}

\begin{document}
\title{On inflation and torsion in cosmology}

\author{Christian G. B\"ohmer\footnote{E-mail: boehmer@hep.itp.tuwien.ac.at}\\
        The Erwin Schr\"odinger International Institute for Mathematical\\ 
        Physics, Boltzmanngasse 9, A-1090 Wien, Austria\\[1ex]
   \and Institut f\"ur Theoretische Physik, Technische Universit\"at Wien,\\
        Wiedner Hauptstrasse 8-10,A-1040 Wien, Austria}
\maketitle

\thispagestyle{fancy}
\setlength{\headrulewidth}{0pt}
\rhead{Preprint ESI 1661, TUW--05--10}

\begin{abstract}
In a recent letter by H.~Davoudiasl, R.~Kitano, T.~Li and H.~Murayama 
``The new Minimal Standard Model'' (NMSM) was constructed that incorporates new 
physics beyond the Minimal Standard Model (MSM) of particle physics. 
The authors follow the principle of minimal particle content and 
therefore adopt the viewpoint of particle 
physicists. It is shown that a generalisation of the geometric
structure of spacetime can also be used to explain physics beyond the MSM.
It is explicitly shown that for example inflation, i.e.~an exponentially expanding
universe, can easily be explained within the framework of Einstein-Cartan theory.
\\ \mbox{} 
\end{abstract}
  
\section{Introduction}

There are many ideas how physics beyond the Minimal Standard Model may be 
explained, however none of them so far was able to give a consistent output
that could explain all experimental results of particle physics and cosmology 
consistently. In contrast to these modern approaches the
authors of~\cite{Davoudiasl:2004be} adopt a conservative particle physicist's point 
of view and include the minimal number of new degrees of freedom to formulate 
the NMSM that can explain Dark Energy, non-baryonic Dark Matter etc.

From a geometrical point of view it may be preferable to allow more general 
geometric structures rather than increasing the number of required particles.
Therefore the guiding principle of this note may be called the principle
of minimal geometry content.

The cosmological principle states that the universe is spatially
homogeneous and isotropic. More mathematically speaking the
four-dimensional (4d) spacetime $(\mathcal{M},g)$ is foliated by 3d
spacelike hypersurfaces of constant time which are the orbits of
a Lie group $G$ acting on $\mathcal{M}$ with isometry group $SO(3)$.
All fields are invariant under the action of $G$. The cosmological
principle implies
\begin{align}
      \lie_{\xi^m} g_{\mu\nu}=0\,,\qquad\mbox{and}\qquad
      \lie_{\xi^m} T^{\lambda}{}_{\mu\nu}=0\,,
      \label{eq:cos1}
\end{align}
where $\xi^m$ are the six Killing vectors (labelled by $m$)
generating the spacetime isometries. $g_{\mu\nu}$ denotes the
metric tensor and $T^{\lambda}{}_{\mu\nu}$ stands for the
torsion tensor, Greek indices label the holonomic components.

By imposing the restrictions~(\ref{eq:cos1}), the metric tensor 
is of Robertson-Walker type
\begin{align}
      ds^2=-dt^2 + \Bigl(\frac{a(t)}{1-\frac{k}{4}r^2}\Bigr)^2 (dx^2+dy^2+dz^2)
          =\eta_{ij}e^i \otimes e^j \,,
      \label{eq:cos2}
\end{align}
where $r^2=x^2+y^2+z^2$ and where the 3-space is spherical for $k=1$,
flat for $k=0$ and hyperbolic for $k=-1$. The vielbein 1-forms
in~(\ref{eq:cos2}) read
\begin{align}
      e^t = dt\,,\quad e^x = \frac{a(t)}{1-\frac{k}{4}r^2} dx\,,
      \quad e^y = \frac{a(t)}{1-\frac{k}{4}r^2} dy \,,
      \quad e^z = \frac{a(t)}{1-\frac{k}{4}r^2} dz \,,
      \label{eq:cos3}
\end{align}
where Latin indices label the anholonomic components.

When the restrictions~(\ref{eq:cos1}) are imposed on the torsion 
tensor~\cite{Tsamparlis:1979}, the (non-vanishing) allowed components are
\begin{align}
      T_{xxt} = T_{yyt} = T_{zzt} = h(t)\,,
      \label{eq:cos4} \\
      T_{xyz} = T_{zxy} = T_{yzx} = f(t)\,,
      \label{eq:cos5}
\end{align}
where we closely follow the notation of~\cite{Goenner:1984}.

\section{Einstein-Cartan theory in cosmology}

In the following it is shown that inflation can be explained without 
introducing additional fields but considering a spacetime with
torsion. The simplest theory of this type is
Einstein-Cartan theory which is derived from the Einstein-Hilbert action
by varying the vielbein and the spin-connection independently. 
Then the field equations are~\cite{Hehl:1976kj}
\begin{align}
      R^i{}_j - \frac{1}{2}R \delta^i_j + \Lambda\delta^i_j 
      = 8\pi\, t^i{}_j \,,
      \label{eq:r} \\
      T^i{}_{jk}-\delta^i_j T^l{}_{lk}-\delta^i_k T^l{}_{jl}
      = 8\pi\, s^i{}_{jk} \,,
      \label{eq:t}
      \end{align}
where $t^i{}_j$ is the canonical energy-momentum tensor and $s^i{}_{jk}$ 
is the tensor of spin. 

By taking the cosmological principle into account the field
equations~(\ref{eq:r}) of Einstein-Cartan theory simplify to
\begin{align}
      3\Bigl(\bigl(h+\frac{\dot{a}}{a}\bigr)^2+\frac{k}{a}-\frac{1}{4}f^2 \Bigr)
      -\Lambda = 8\pi\rho\,,
      \label{eq:cos10} \\
      -\Bigl(\bigl(h+\frac{\dot{a}}{a}\bigr)^2+\frac{k}{a}-\frac{1}{4}f^2 \Bigr)
      -2\Bigl(\bigl(h+\frac{\dot{a}}{a}\bigr)^{\dot{}}
      +\frac{\dot{a}}{a}\bigl(h+\frac{\dot{a}}{a}\bigr)\Bigr)
      +\Lambda = 8\pi P\,.
      \label{eq:cos11}
\end{align}
The torsion field equations~(\ref{eq:t}) become
\begin{alignat}{2}
      f &= 8\pi s\,, &\qquad s(t)&=S_{xyz}=S_{zxy}=S_{yzx}\,,
      \label{eq:cos12} \\
      -2h &= 8\pi q\,, &\qquad q(t)&=S_{xxt}=S_{yyt}=S_{zzt}\,.
      \label{eq:cos13}
\end{alignat}
If no torsion source is present $s=q=0$, the algebraic equations of
motion imply the vanishing of the torsion tensor $f=h=0$. 
Without torsion, the field equations~(\ref{eq:cos10})
and~(\ref{eq:cos11}) reduce to the standard Friedman equations of
cosmology.

Let us have a closer look at the field equations~(\ref{eq:cos10})--(\ref{eq:cos13})
in case of $q=h=0$, i.e. only the skew-symmetric part of the torsion tensor,
cf~\cite{Minkowski:1986kv}. Then the field equations simplify to
\begin{align}
      3\Bigl( \bigl(\frac{\dot{a}}{a}\bigr)^2 +\frac{k}{a}-\frac{1}{4}f^2\Bigr)
      -\Lambda &= 8\pi\rho \,, 
      \label{eq:cos14}\\
      -\Bigl( \bigl(\frac{\dot{a}}{a}\bigr)^2 +\frac{k}{a}-\frac{1}{4}f^2\Bigr)
      -2\Bigl( \bigl(\frac{\dot{a}}{a}\bigr)^{\dot{}} +\bigl(\frac{\dot{a}}{a}\bigr)^2 \Bigr)
      +\Lambda &= 8\pi P \,, 
      \label{eq:cos15}\\
      f &= 8\pi s \,, 
      \label{eq:cos16}
\end{align}
which implies the following conservation equation
\begin{align} 
      \frac{\dot{\rho}}{3} + \frac{\dot{a}}{a}(\rho+P)
      +\frac{s}{2}(\dot{f}+\frac{\dot{a}}{a}f) =0 \,.
      \label{eq:cos17}
\end{align}
With~(\ref{eq:cos16}) the two remaining independent field equations
can be reformulated to give
\begin{align}
      3\Bigl( \bigl(\frac{\dot{a}}{a}\bigr)^2 +\frac{k}{a}\Bigr)
      &=8\pi \rho_{{\rm eff}} = 8\pi\rho + \Lambda +\frac{3}{4}(8\pi s)^2\,,
      \label{reff} \\
      -2\frac{\ddot{a}}{a}-\bigl(\frac{\dot{a}}{a}\bigr)^2-\frac{k}{a}
      &=8\pi P_{{\rm eff}} = 8\pi P - \Lambda -\frac{1}{4}(8\pi s)^2\,,
      \label{peff}
\end{align}

In~(\ref{reff}) and~(\ref{peff}) the matter dominated era of cosmology is 
defined by $P=0$ and $\rho=\rho_{{\rm m}}$ where in addition it is
assumed that the torsion contribution is sufficiently small, which is
indeed very reasonable as shall be seen. The radiation dominated era
is defined by the equation of state $P=\rho/3$ and $\rho=\rho_{{\rm r}}$, again
with an sufficiently small torsion contribution. For sake of simplicity
we assume the following setup for the {\em torsion dominated era}, in which
the universe is exponentially increasing: Assume that torsion in~(\ref{reff})
and~(\ref{peff}) is the leading contribution, such that one may neglect
the others. In the early time of the universe the particle density was high
and therefore the probability of having some non-vanishing macroscopic spin
is the higher the denser the matter distribution is. On the 
other hand it is reasonable that the averaged spin density is 
exponentially decreasing decreasing with time, $s\propto \exp(-t/\tau)$,
where $\tau$ is a characteristic time scale. 
Putting this into~(\ref{eq:cos17}) yields
\begin{align}
      \frac{\dot{s}}{s}=-\frac{1}{\tau}=-\frac{\dot{a}}{a}\,,
\end{align}
which simply implies that the scale factor $a$ is an exponentially
increasing function of time, $a\propto\exp(t/\tau)$ if the torsion function 
is exponentially decreasing and if the torsion contribution is 
the leading one.

Hence a physically intuitive assumption on the behaviour of torsion
can explain the inflation era of cosmology without introducing 
further particles. Since the torsion is rapidly decreasing, its
contribution to~(\ref{reff}) and~(\ref{peff}) will indeed be sufficiently small
after the short period of inflation. This implies that todays
cosmological measurements possibly should detect some small non-vanishing 
torsion contribution, (see e.g~\cite{GarciadeAndrade:1999qt}). This torsion 
remnant could then be used to solve the sign problem of the cosmological 
constant, as was shown by the author in~\cite{Boehmer:2003iv}.

It is neither the author's aim to criticise the motivation and
derivation of the NMSM nor to criticise the successful way
that lead to the MSM. We try to show that other, 
equally conservative, approaches may also work. It should be emphasised 
that the consideration of torsion is nearly as old as general 
relativity itself (see e.~g.~\cite{Goenner:2004se} for a historical review).
Thus the guiding principle of minimal geometry content might
be as successful as the minimal particle content principle.
Only the experiment will decide which of these two principles 
is the one describing nature correctly.

\section*{Acknowledgements}

I wish to thank Herbert Balasin and Wolfgang Kummer for valuable 
comments. Moreover I wish to thank Dominik~J.~Schwarz for the 
useful discussion.

The work was supported by the Junior Research Fellowship of The
Erwin Schr\"odinger International Institute for Mathematical Physics.

\end{document}